\documentclass[12pt,preprint]{aastex}

\shortauthors{Weinberger et al.}
\shorttitle{TW Hya Disk}

\begin{document}
\renewcommand{\topfraction}{0.99}
\renewcommand{\bottomfraction}{0.99}
\renewcommand{\floatpagefraction}{0}

\title{Infrared Views of the TW Hya Disk}

\author{A. J. Weinberger\altaffilmark{1,2}, E. E. Becklin\altaffilmark{1},
G. Schneider\altaffilmark{3}, E. I. Chiang\altaffilmark{4},
P. J. Lowrance\altaffilmark{1}, M. Silverstone\altaffilmark{3},
B. Zuckerman\altaffilmark{1}, D. C. Hines\altaffilmark{3}, and
B. A. Smith\altaffilmark{5}}

\altaffiltext{1}{Division of Astronomy, UCLA, Box 951562,
Los Angeles, CA 90095-1562, becklin,lowrance,ben@astro.ucla.edu}

\altaffiltext{2}{Department of Terrestrial Magnetism, Carnegie
Institution of Washington, 5241 Broad Branch Rd. NW, Washington, DC
20015, weinberger@dtm.ciw.edu}

\altaffiltext{3}{Steward Observatory, University of Arizona, 933
N. Cherry Ave., Tucson, AZ 85721, gschneider,murray,dhines@as.arizona.edu}

\altaffiltext{4}{Department of Astronomy, University of California at
Berkeley, 601 Campbell Hall, Berkeley, CA 94720, echiang@astron.berkeley.edu}

\altaffiltext{5}{Institute for Astronomy, University of Hawaii,
Honolulu, HI, brad@mahina.hawaii.edu}

\begin{abstract}
The face-on disk around TW Hya is imaged in scattered light at 1.1 and
1.6 $\mu$m using the coronagraph in the Near Infrared Camera and Multi
Object Spectrometer aboard the Hubble Space Telescope.  Stellar light
scattered from the optically thick dust disk is seen from 20--230
AU. The surface brightness declines as a power law of r$^{-2.6\pm0.1}$
between 45 and 150 AU.  The scattering profile indicates that the disk
is flared, not geometrically flat.  The disk, while spatially unresolved
in thermal radiation at 12 and 18 $\mu$m in observations from the
W. M. Keck Observatory, shows amorphous and crystalline silicate
emission in its spectrum.  A disk with silicate grains of a $\sim$1
micron in size in its surface layers can explain the color of the
scattered light and the shape of the mid-infrared spectrum.  Much larger
grains in the disk interior are necessary to fit the millimeter-wave
spectral energy distribution, and hence grain growth from an original
interstellar size population may have occurred.
\end{abstract}

\keywords{stars:circumstellar matter,individual(TW Hya) -- infrared
radiation}

\section{Introduction}

TW Hya (= CD--34 7151, H = 7.6 mag) is a star with a large H$\alpha$
emission equivalent width, visual variability and location at 23$^\circ$
galactic latitude, and it was thus labeled as ``peculiar'' by
\citet{Henize}. \citet{Rucinski} identified it as a spectral type K7Ve T
Tauri star far from any molecular cloud (i.e. ``isolated'').  TW Hya
displays classical signs of being surrounded by an accretion disk,
including variable H$\alpha$ emission \citep{Muzerolle} and excess flux
over photospheric in the ultraviolet and near-infrared.  In addition it
has disk signatures in the form of far-infrared \citep{Rucinski85},
submillimeter continuum \citep{Weintraub89} and CO line emission
\citep{Zuckerman95}. All of these disk indicators are large for a star
as old as TW Hya, which has an age of $\sim$8 Myr determined from
pre-main sequence evolutionary tracks \citep{Webb99}. Zuckerman et al.
inferred a face-on viewing geometry for the disk based on the small CO
line widths they observed.  Recent imaging observations do indeed reveal
a circular disk. Scattered radiation from the disk has been imaged at
visible and near-infrared wavelengths \citep{Krist00,Trilling01}, and
circumstellar dust emission has been imaged at a wavelength of 7 mm
using the VLA \citep{Wilner00}.

The Hipparcos mission measured the distance to TW Hya as 56$\pm$7 pc
\citep{Wichmann98}. An eponymous association of $\sim$20 star systems
has been identified from x-ray and proper motion surveys within
10$^\circ$ and 10 pc of TW Hya
\citep{delareza,Webb99,sterzik99,Zuckerman01}.  The approximately coeval
members of the TW Hya Association are amongst the products of recent
star formation closest to the Sun.  However, only three other members
have measurable far-infrared excesses from IRAS: HR 4796A, HD 98800, and
Hen3-600 (L$_{IR}$/L$_\star$=0.01, 0.19, and 0.21, respectively). The
optical depth around TW Hya (L$_{IR}$/L$_\star$=0.25) is the
largest and corresponds to the reprocessing expected for a optically
thick disk \citep{Adams87}.

The state of disks of intermediate age ($\sim$10$^7$ yr) has not been
well studied.  The structure of the youngest disks ($<$10$^6$ yr),
hundreds to thousands of AU in radius around embedded protostars have
been copiously studied with millimeter interferometry
\cite[e.g.]{beckwith} and visual and infrared imaging
\cite[e.g.]{burrows96,padgett}.  There are only a few examples of older
late-type stars surrounded by disks which have been spatially
resolved in the near-infrared, most of which are in binary systems
\cite[e.g.]{roddier96,koresko}.  At ages $\gtrsim$10$^7$ yr, most known
disks orbit early type stars because only in these systems do the disks
scatter enough photons to be detectable (e.g. $\beta$ Pic
\citep{smith84}).  TW Hya, thanks to its proximity to the Sun and its
age, is thus an excellent laboratory for studying a disk of intermediate
age around a low mass star.  This age range is thought to be critical in
the evolution of the planetesimals that form the cores of gas giant and
terrestrial planets \citep{Weidenschilling}.

\section{Observations and Data Analysis}
\subsection{NICMOS}\label{analysis}

TW Hya was observed at two different wavelength bands with the Near
Infrared Camera and Multi Object Spectrometer (NICMOS) aboard the Hubble
Space Telescope.  The star was placed in the coronagraphic hole in
camera 2 ($\sim$0.076 \arcsec\ pixel$^{-1}$, $\sim$19\arcsec\ square
field of view).  The observational strategy was the same for both -- a
total of 1216 s of integration taken in three non-destructive
multiple-readout (MULTIACCUM) sequences at each of two telescope
orientations, rolled 7$^\circ$ with respect to each other about the
target axis.  To minimize time-dependent point spread function (PSF)
variations, the rolled image sets were obtained within $\sim$40 minutes
in a single target visibility period.  The first observation was made
with the F160W filter ($\lambda_{central}$=1.59\micron,
FWHM=0.40\micron) on 1998 August 16 and the second was made with the
F110W filter ($\lambda_{central}$=1.10\micron, FWHM=0.59\micron) on 1998
November 25.  A summary of the observations is presented in Table
\ref{tab_nicmosobs}.

Prior to each set of coronagraphic observations, two 0.86~s target
acquisition frames were obtained with the F171M
($\lambda_{central}$=1.72 \micron, FWHM=0.07 \micron) filter.
Contemporaneous lamp flats and backgrounds were obtained at F160W for
the purpose of locating the coronagraphic hole and enabling good
flat-fielding near the hole. Special coronagraphic flat fields at each
band were created by modifying a standard high signal to noise ratio
on-orbit reference flat field. In a 16$\times$16 pixel region around and
including the coronagraphic hole, the pixel sensitivities were measured in the
F160W internal flat field (hole-finding) image, scaled to F110W if
necessary, and replaced in the reference flat.

\begin{deluxetable}{lllllll}
\tablecaption{Log of NICMOS Observations in Programs 7226 and 7233\label{tab_nicmosobs}}
\tablewidth{0pt}
\tablehead{
\colhead{Filter} &\colhead{TW Hya visit}
     &\colhead{Date} &\colhead{Time (UT)} &
                   \colhead{PSF Star} &
                          \colhead{Visit} &
                                \colhead{Date}}
\startdata
F160W  &7  &1998 Aug 16 &4:05--4:19   & HR 8721 & 81\tablenotemark{a} & 1998 Aug 11\\
F160W  &8  &1998 Aug 16 &4:35--4:48   & HR 8721 & 81\tablenotemark{a} & 1998 Aug 11\\
F110W  &91 &1998 Nov 25 &20:15--20:27 &$\tau^1$ Eri & 80              & 1998 Nov 14\\
F110W  &92 &1998 Nov 25 &20:42--20:55 &$\tau^1$ Eri & 81              & 1998 Nov 14\\
\enddata

\tablenotetext{a}{HR 8721 drifts behind the coronagraph in this visit due
to the loss of lock on one guide star.  We used only the first 240 s of
integration in which the drift was minimal. In this time, the SNR of HR 8721
is 1.5 times that of TW Hya.}
\end{deluxetable}

The MULTIACCUM data sets were processed with ``nicred''
\citep{MacLeod97}.  Dark frames prepared by the NICMOS instrument
definition team were used to subtract the dark current and correct the
detector shading. After bias subtraction, linearization and flat
fielding, known bad pixels were replaced with a Gaussian weighted
average of neighboring pixels, and the three MULTIACCUM images from each
spacecraft orientation were medianed.  The best available photometric
calibration was applied to the final images (F110W: 1 ADU s$^{-1}$ =
2.031$\times$10$^{-6}$ Jy and, referenced to Vega, 0~mag = 1775 Jy;
F160W: 1 ADU s$^{-1}$ = 2.190$\times$10$^{-6}$ Jy and 0~mag = 1083 Jy).
The uncertainty in this absolute photometric calibration is $\sim$2\%
(M. Rieke, private communication).

In the reduced and calibrated NICMOS images, scattered and diffracted
light from the star can be much higher than the disk flux, so the
instrumental PSF from an unresolved source must be subtracted.  The same
observing strategy, including the telescope roll, was used in our GTO
program for 14 other stars at F110W and 77 at F160W.  These were all
processed in the same way as the TW Hya images, and we use targets which
do not show extended disk emission as a library of possible PSF stars.
The detailed character of the NICMOS PSF changes with time primarily due
to thermal variations in the HST optical assembly \citep{kulkarni00} and
small shifts of the camera cold mask \citep{Krist98}; hence some PSF
star subtractions produce much lower residuals than others
\citep{Schneider01}.  The software IDP3 \citep{Lytle} utilizing cubic
convolution interpolation was used to register coronagraphic PSFs to
both orientations of the TW Hya observations.  Only stars brighter than
TW Hya were considered, so the signal to noise ratio in the PSFs were
greater than that of TW Hya at every radius, and the PSF subtraction
residuals were dominated by systematic uncertainties.  To minimize the
effect of the cold mask shift, we considered PSFs taken within one month
of TW Hya.  Five F160W PSFs and one F110W PSF met all criteria.

\begin{figure}
\plotone{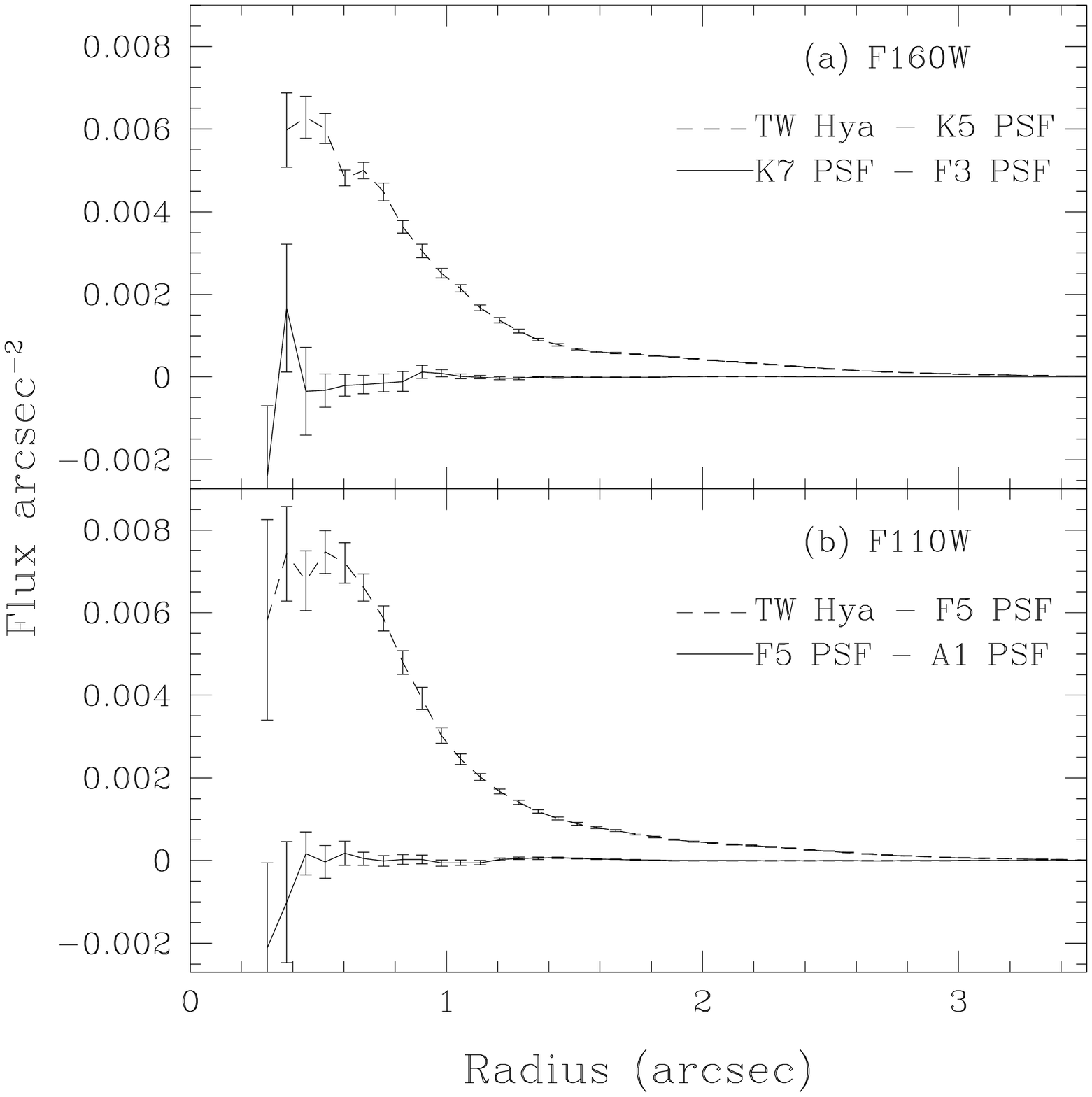}
\figcaption[f1.eps]{(a) Comparison of TW Hya with PSF null at
F160W. The spectral types of TW Hya (K7) and its PSF (GL 879; K5) are
quite close.  To illustrate the effect of subracting a PSF of a
different spectral type, the null is the result of subtracting an
F3-type PSF from a K7-type PSF.  (b) Comparison of TW Hya and a PSF Null
at F110W.  The spectral type of the PSF, $\tau^1$ Eri is F5, so color
residuals should not be greater than the level demonstrated in the PSF
null of part {\it a}.  The PSF null in this case is of $\tau^1$ Eri
minus an A1-type PSF.  For all these subtractions, comparions with Tiny
Tim show that the residuals are dominated by time-variable effects and not
by color mismatch. \label{fig_psfcompare}}.
\end{figure}

The fine structure of the broad band PSF is affected by the spectral
energy distribution of the source.  When subtracting a reference point
source, therefore, mismatches in effective temperature can result in
increases in the subtraction residuals.  At F160W, where several
possible PSF observations existed, the chosen PSF is the K4V star HR
8721 (=GL 879) which is quite close to TW Hya (K7) in spectral type.  At
F110W, however, where many fewer PSFs were taken, the spectral type of
the chosen PSF, $\tau^1$ Eridani, is F4V.  We tested that the effect of
the color mismatch is small in two ways.  First, we subtracted
coronagraphic PSFs of different colors and examined the residuals; two
examples are plotted in Figure \ref{fig_psfcompare}.  Although color
mismatch does increase the residuals, it does not introduce a false
excess that could mimic a disk, nor does it significantly affect the
shape of the observed TW Hya disk.  Second, we created synthetic PSFs of
different spectral types with Tiny Tim \citep{KristTT}.  Since Tiny Tim
models diffraction and not scattering from the coronagraph or other
elements in the optical path, its PSFs isolate the effects of color
alone.  The Tiny Tim subtraction residuals were significant only within
0.6\arcsec\ of the star and were smaller than those from real PSF
subtractions.

The dominant source of photometric uncertainty in the PSF-subtracted
images results from uncertainty in how to appropriately scale the PSF to
the brightness of TW Hya.  There are two possible methods to determine
the scaling factor: a) the use of a priori photometry of the two stars
and b) the use of measurements made on the subtracted image, including
nulling the flux in the diffraction spikes and demanding that the flux
approach zero at large radii from the star.  The former suffers from the
fact that no images of the stars are taken in the broad-band filters
used for the coronagraphic imaging, because the stars saturate the
detector in the minimum integration time, so stellar models must be
employed for the color transformations from the acquisition filters.
The latter suffers from the fact that the disk contaminates the
measurements.

Our PSF scalings were determined by minimizing the flux in the
subtracted diffraction spikes outside the region of apparent disk flux.
We scale the PSF images in ADU s$^{-1}$ for $\tau^1$ Eri (F110W) by
0.0099 and Gl 879 (F160W) by 0.031. The uncertainty in this scaling is
3\% based on the variation between the four diffraction spikes. This
uncertainty in the scaling results in approximately a 10\% uncertainty
in the disk photometry in each band.

\subsection{Keck}

Images and spectra of TW Hya were taken with the facility Long
Wavelength Spectrograph (LWS) \citep{jones93} on the 10~m Keck I
telescope on UT 11 December 2000 and 4--5 February 2001. During all
three nights, the weather was photometric with low water vapor optical
depth. The images were limited by seeing with FWHM =0.$''$57 and
0.$''$50 at 11.7 \micron\ and 17.9 \micron\ respectively.  LWS uses a
$128\times128$\ pixel Boeing Si:As detector, and has a plate scale of
0.08 arcsec pixel$^{-1}$, resulting in a focal- plane field of view of
$10.24\arcsec$ square.  It also provides a dispersion of 0.037 $\mu$m
pixel$^{-1}$; when combined with a 6 pixel (0.$''$48) slit, the resolution
is R $\sim$ 120.  When viewed through the long slit, vignetting limits
the field of view to 8.5\arcsec. Observational details are found in
Table \ref{tab_keckobs}.

 In spectroscopic mode, data were taken with the telescope secondary
chopped in the same direction as the nod with a frequency of 5 Hz and
amplitude of 10$''$. Only the on-source chop fell on the detector.  In
preliminary data analysis, the images were double differenced to
subtract the thermal background and bad pixels were corrected by
interpolation.  The infrared bright star HR 4532 (11.7 $\mu$m flux
density = 53.6 Jy) was observed just after TW Hya to provide atmospheric
calibration.

In imaging mode, data were obtained by chopping the secondary at 5 Hz
and nodding after $\sim$20~s.  In some cases, the data were taken by
chopping and nodding 5'', so four images appear on the detector; in
others the throw was 10'' so one image appears on the detector.  In
either case, every double differenced (i.e. fully background subtracted)
image was centroided and recentered to remove telescope drift.

\begin{deluxetable}{llcclcc}
\tablecaption{Log of Mid-Infrared Observations\label{tab_keckobs}}
\tablewidth{0pt}
\tablehead{
\colhead{Mode}&\colhead{Object\tablenotemark{1}}
                      &\colhead{Filter center}
                            &\colhead{Filter width}
			           &\colhead{Date}
                                                &\colhead{Integration}
                                                     &\colhead{Airmass}\\
\colhead{}    &\colhead{}
		      &\colhead{($\mu$m)}
		            &\colhead{($\mu$m)}
                                  &\colhead{}
                                                &\colhead{Time (s)}
                                                     &\colhead{}
}
\startdata
Imaging       &TW Hya &11.7 &1.0  &11 Dec 2000  &120 &1.7 \\
	      &TW Hya &11.7 &1.0  &04 Feb 2001   &216 &1.7 \\
              &PSF    &11.7 &1.0  &04 Feb 2001   &216 &1.6 \\
	      &TW Hya &17.9 &2.0  &05 Feb 2001   &809 &1.8 \\
              &PSF    &17.9 &2.0  &04 Feb 2001   &216 &1.8 \\
Spectroscopy  &TW Hya &10.5 &4.9  &11 Dec 2000  &408 &1.7 \\
              &Atmos. Cal.    &10.5 &4.9  &11 Dec 2000  &96  &1.5 \\
\enddata
\tablenotetext{1}{HR 4532 was used as the PSF and atmospheric
calibrator.  The bright star HR 3748 (11.7 \micron\ flux density = 53.6
Jy) was observed for photometric calibration.}
\end{deluxetable}

The spectra of TW Hya and of the atmospheric calibrator were extracted
by fitting Gaussians in the spatial direction to the flux at every
spectral pixel. The uncertainty in the flux at every wavelength was
estimated from the noise in the sky on either spatial side of the long
slit image. The location of the Telluric O$_3$ line at $\sim$9.6\micron\
was measured in the sky spectra associated with both TW Hya and the
standard, from which it was determined that small unrepeatabilities in
the grating locator mechanism introduced a small shift in the central
wavelength between the two.  Before division of the TW Hya spectrum by
the standard spectrum, the standard spectrum was shifted in wavelength
by 2.5 pixels to match that of TW Hya.  After division, the resultant
spectrum was multiplied by a blackbody at the temperature of the
standard star (3100 K) in order to recover the intrinsic flux
distribution.  The location of the Telluric ozone line was measured from
a cross correlation with atmospheric data from the National Solar
Observatory sunspot atlas \citep{NSOspot} and used for wavelength
calibration.

\section{Results} \label{results}
\subsection{Reflected Light}

\begin{figure}
\plotone{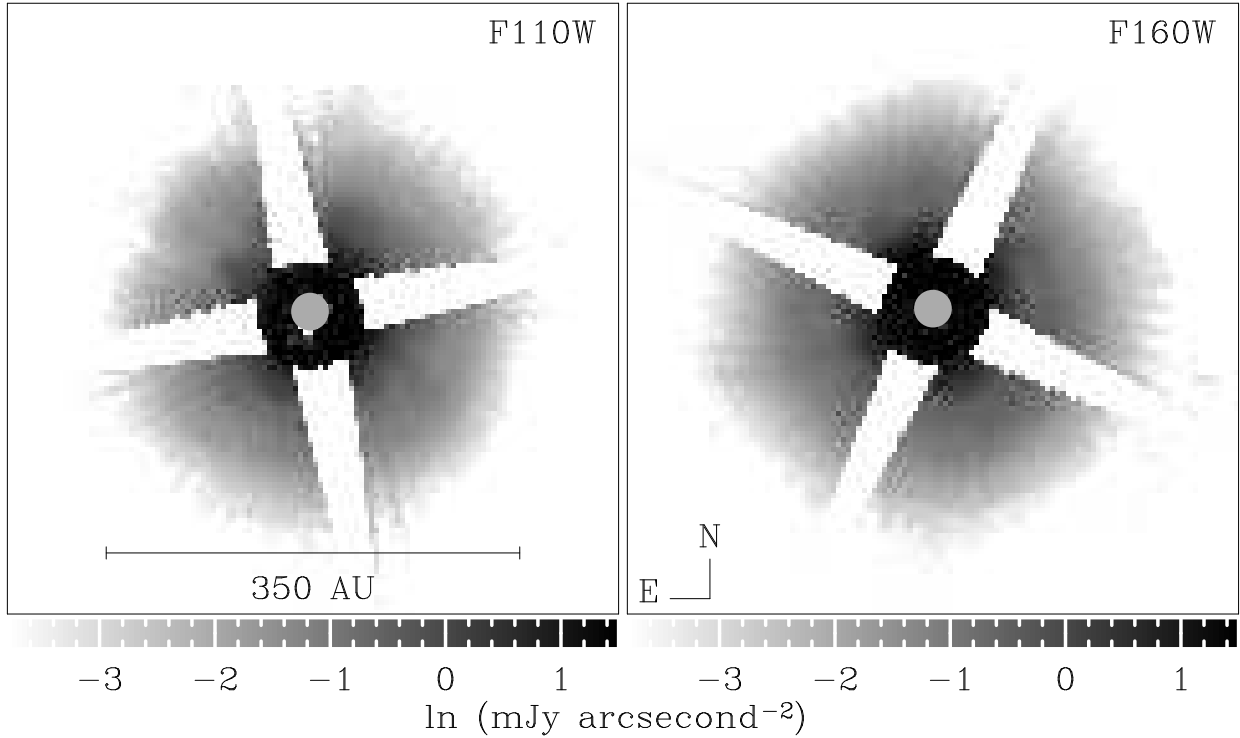} \figcaption[f2.eps]{{\it Left:} PSF subtracted,
roll-combined image of TW Hya at F110W; {\it Right:} PSF subtracted,
roll-combined image of TW Hya at F160W, both shown with a natural
logarithmic stretch and on the same spatial scale and orientation.  This
stretch is employed since the disk falls off as a power law and so it
highlights the outer disk. The size (radius 0.$''$3) and location of the
coronagraph are shown with the central gray circle.  The diffraction
spikes have been masked in regions where their noise is greater than the
disk signal.  They have a tapered shape because pixels obscured at one
telescope orientation were replaced with those not obscured at the other
angle.\label{fig_images}}
\end{figure}

The NICMOS PSF subtracted images of TW Hya at F110W (1.1\micron) and
F160W (1.6\micron) are shown on a natural logarithmic stretch in Figure
\ref{fig_images}.  The disk becomes visible
just outside the coronagraph and continues out to a radius of 4$''$
($\sim$230 AU).  The noise computed in the background of the images is
21.1 and 21.2 mag arcsec$^{-2}$ at F160W and F110W respectively.  In the
disk, the noise is dominated by subtraction residuals.  The spatial
resolution of these images is 0.$''$15 at F160W and 0.$''$12 at F110W.

The diffraction spikes from the stellar cores are particularly sensitive
to time-dependent instrument and telescope variations and never subtract
out perfectly.  Where possible, the pixels corrupted by the spikes in
images from one telescope orientation were replaced by uncorrupted
pixels from the other orientation.  This results in the tapered
appearance of the spikes in the final images.  The spikes were not
masked in the central 0.$''$9, because even though they are quite noisy,
the very bright inner disk of TW Hya can still be significantly detected
above them in the region from 0.$''$38 (just outside the edge of the
coronagraphic hole) to 0.$''$9.

The total disk flux densities were measured in an annulus between radii
of 0.$''$38 and 4$''$ (22$-$228 AU) to be 17.4$\pm$1.8 mJy (12.5 $\pm$
0.1 mag) at F110W and 21.6$\pm$ 2.2 mJy (11.7 $\pm$ 0.1 mag) at
F160W. At both wavelengths, the pixels masked by the diffraction spikes
were replaced by the average value of the flux density at their radii
and included in the photometry reported above.  The quoted uncertainties
are not dominated by photon counting statistics but rather by the
systematic uncertainty in how to scale the PSF stars to TW Hya.  The
measured ratios of scattered to stellar light are 0.024 and 0.021 at
F110W and F160W respectively.  The surface brightness peaks at a radius
of 0.$''$5 (29 AU) at 4.5$\pm$1.2 mJy arcsec$^{-2}$ at F110W and
5.7$\pm$1.4 mJy arcsec$^{-2}$ at F160W and declines smoothly with
radius.

To measure any ellipticity in the disk, the radii of seven independent
isophotal contours were calculated as a function of azimuthal angle in
the F160W image.  For a more significant result, the contours
were normalized to a radius of one and averaged in sixteen azimuthal
bins.  The average isophote was fit with an ellipse using eccentricity
(i.e. inclination) and position angle as two free parameters.  The
resulting best fit had e=0.025$\pm$ 0.014.  If this ellipticity is
interpreted as due to the inclination of an intrinsically circular disk,
the measured inclination is of marginal significance, but a robust
3$\sigma$ upper limit to the inclination is 4$^\circ$ from face-on.

\begin{figure}
\plotone{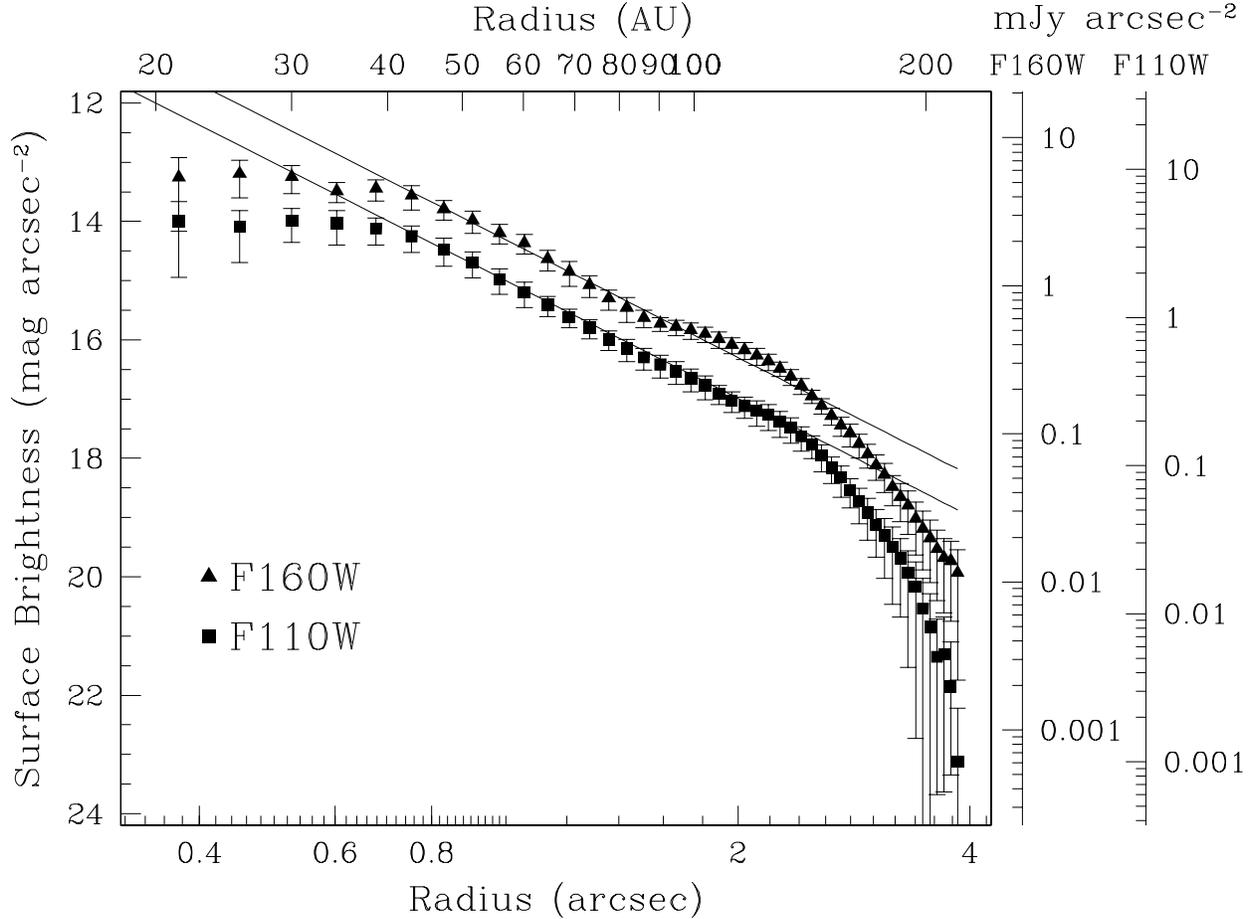}
\figcaption[f3.eps]{Surface brightness of TW Hya at both 1.1\micron\ and
1.6\micron\ as a function of radius.  The black lines show a power law
of r$^{-2.6}$ fit between 40 and 150 AU and extrapolated to other
radii. The error bars on each point represent the standard deviation of
all the pixels at the given radius in an annulus one pixel
wide. \label{fig_magsurf}}
\end{figure}

The azimuthally averaged disk surface brightnesses at both wavelengths
are shown in Figure \ref{fig_magsurf}.  Between radii of 40 AU and 150
AU, the disk at both wavelengths is well fit by a single power law,
r$^{-2.6 \pm 0.1}$ (the solid lines in Figure \ref{fig_magsurf}).
However, also seen at F160W is a ``wiggle'' in the surface brightness
centered at 1.$''$5 (85 AU).  Beyond 150 AU, the disk falls off more
sharply. Within 40 AU, it appears to flatten.

To obtain the color of the disk, the images at F110W and F160W were
ratioed.  Since the images in the two filters were taken at quite
different telescope orientations, the diffraction spikes corrupt 50\% of
the ratioed image.  There are no significant gradients in the color as a
function of radius. The F110W - F160W color averaged over the pixels
free from diffraction spike contamination is 0.77 $\pm$0.10 mag.

Since TW Hya was never observed with the NICMOS F110W and F160W filters
without the coronagraph, the color of the star at these two NICMOS bands
was estimated based on a stellar photosphere model by \citet{kurucz}.
The F171M magnitude of the star was measured from the acquisition images
to be 7.39 $\pm$ 0.04 mag from both the August and November 1998 data.
The UBVRI variability of TW Hya, which amounts to $\sim$0.5 mag at V and
decreases toward the red, has been examined in detail and shown to have
an overall period of 1--2 days with much shorter episodes of rapid
change \citep{Rucinski,Mekkaden,Herbst}.  The star becomes bluer by up
to 20\% as it brightens.  The near-infrared variability has not been
thoroughly studied, but its level is expected to be low if the
underlying cause of the variation is from hot-spots.  The high level of
consistency in the photometry of our acquisition images from the two
dates implies that the star did not vary at 1.7 $\mu$m between August
and November.  We assume the star was similarly stable at 1.1 and
1.6\micron.

A Kurucz stellar atmosphere model with T$_{eff}$=3925~K, log(g)=4, and
log(Z)=0 was used to extrapolate the F171M magnitude to F110W and F160W
giving 8.47 and 7.55 mag respectively.  The same model gives H=7.53 and
J=8.35 mag (in the CIT system) which are 10\% brighter than the values
given by \citet{Rucinski}, which had an uncertainty of 5--10\%.  The
inferred J-H color, 0.82, is identical to that obtained by
\citet{Rucinski}, so we believe the model extrapolation to the NICMOS
filters to be accurate.  The F110W$-$F160W stellar color is thus 0.92
mag, while the color of the disk is 0.77$\pm$0.10 mag, so the disk color
appears to be marginally (1.3$\sigma$) blue.

Using HST/WFPC2, \citet{Krist00}, also found a slightly blue color for
the disk at visual wavelengths (0.6 and 0.8 $\mu$m) in images made
when the star was as bright as it has ever been recorded.  The best estimate
of color should come from the largest wavelength range,
i.e. 0.6--1.6~\micron.  Over the disk region between 0.5 and 3$''$,
F606W$-$F160W=2.6 $\pm$ 0.2.  For TW Hya, F606W$-$F160W=2.5, so the disk
appears to scatter neutrally over this 1~\micron\ wavelength baseline.

\subsection{Thermal Emission}

\begin{figure}

\plotone{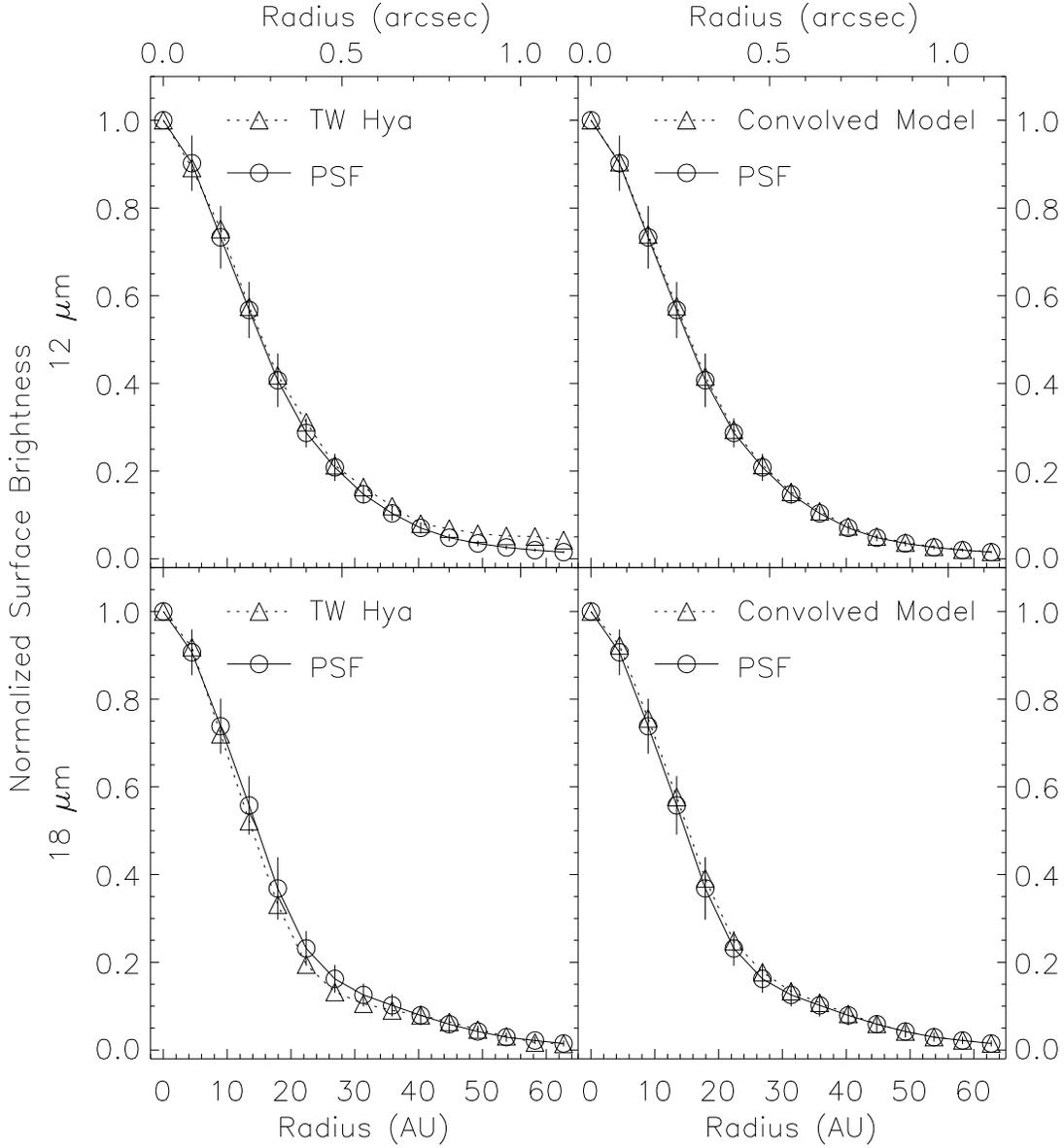} \figcaption{In the left panels, the azimuthal profiles
of our Keck point spread function star 11.7 (top row) and 17.9~\micron\
(bottom row) are compared to profiles of our images of TW Hya.  In the
right panels, the model described in \S\ref{midirsection} convolved with
our PSF is compared to the PSF.  At each radius, the error bars show the
standard deviation about the mean value of pixels in a one pixel wide
annulus.  The PSF is evidently oversampled at these wavelengths, and the
first Airy minimum can just be seen in the 17.9 \micron\ profile at
0.$''$48.  TW Hya is unresolved (left). This result is consistent with
the model (right).\label{fig_midirimages}}
\end{figure}

\begin{figure}
\plotone{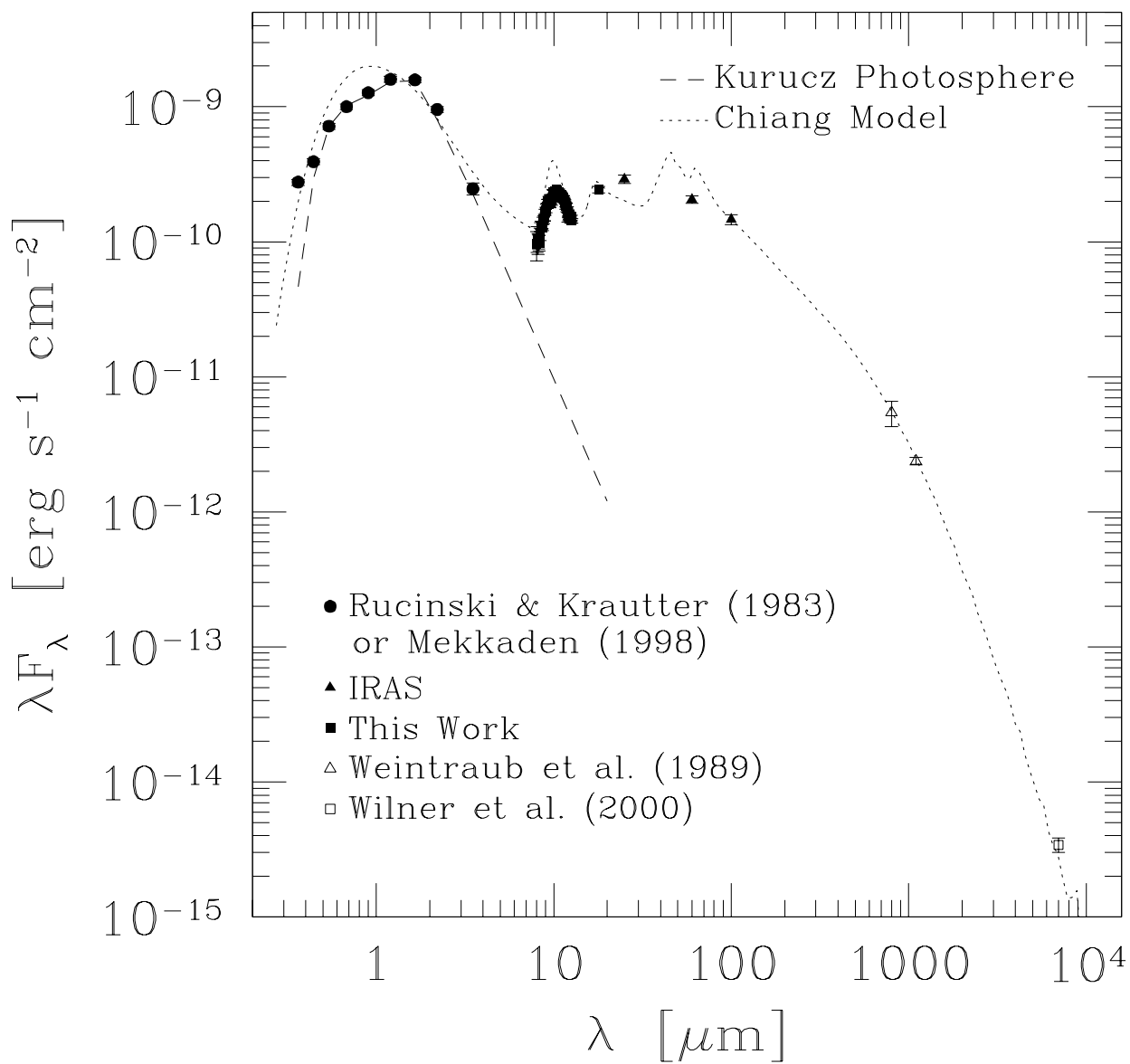} \figcaption[f5e.ps]{Measured spectral energy
distribution for TW Hya compared to that of a Kurucz model stellar
photosphere with T$_{eff}$=3925 K, log(g)=4.0, and Solar metallicity
(dashed line) and the \S\ref{midirsection} disk model (dotted line).
There is no thermal excess at wavelengths $\leq$3\micron, so dust does
not extend to within 0.05 AU. \label{fig_sed}}
\end{figure}

The radial profiles of TW Hya in thermal emission at 11.7 and 17.9
$\mu$m are shown in Figure \ref{fig_midirimages} compared with images of
the PSF star.  TW Hya appears to be the same size as the PSF,
i.e. unresolved, at both wavelengths. Integrated flux densities were
measured in a synthetic aperture of radius 2.$''$4, and were 0.72 $\pm$
0.04 Jy at 11.7\micron\ and 1.45 $\pm$ 0.08 Jy at 17.9\micron. The
measurement uncertainties reflect both statistical and photometric
calibration contributions.  These flux densities agree very well with
those given in the \citet{IRASFSC} at 12 and 25 \micron\ interpolated to
our wavelengths of observation.  Thus, all of the mid-infrared flux in
the IRAS beam arises from quite close to the star. Using a Kurucz model
of a K7 photosphere (see Figure \ref{fig_sed}) to predict the L-band
flux and then assuming Rayleigh-Jeans fall-off from 3.5 to 18 \micron,
we determine the stellar flux density at 11.7 and 18.9 \micron\ as 31
and 13 mJy, respectively.  The mid-infrared color temperature of the
disk is 214 K from our measurements.

\begin{figure}
\plotone{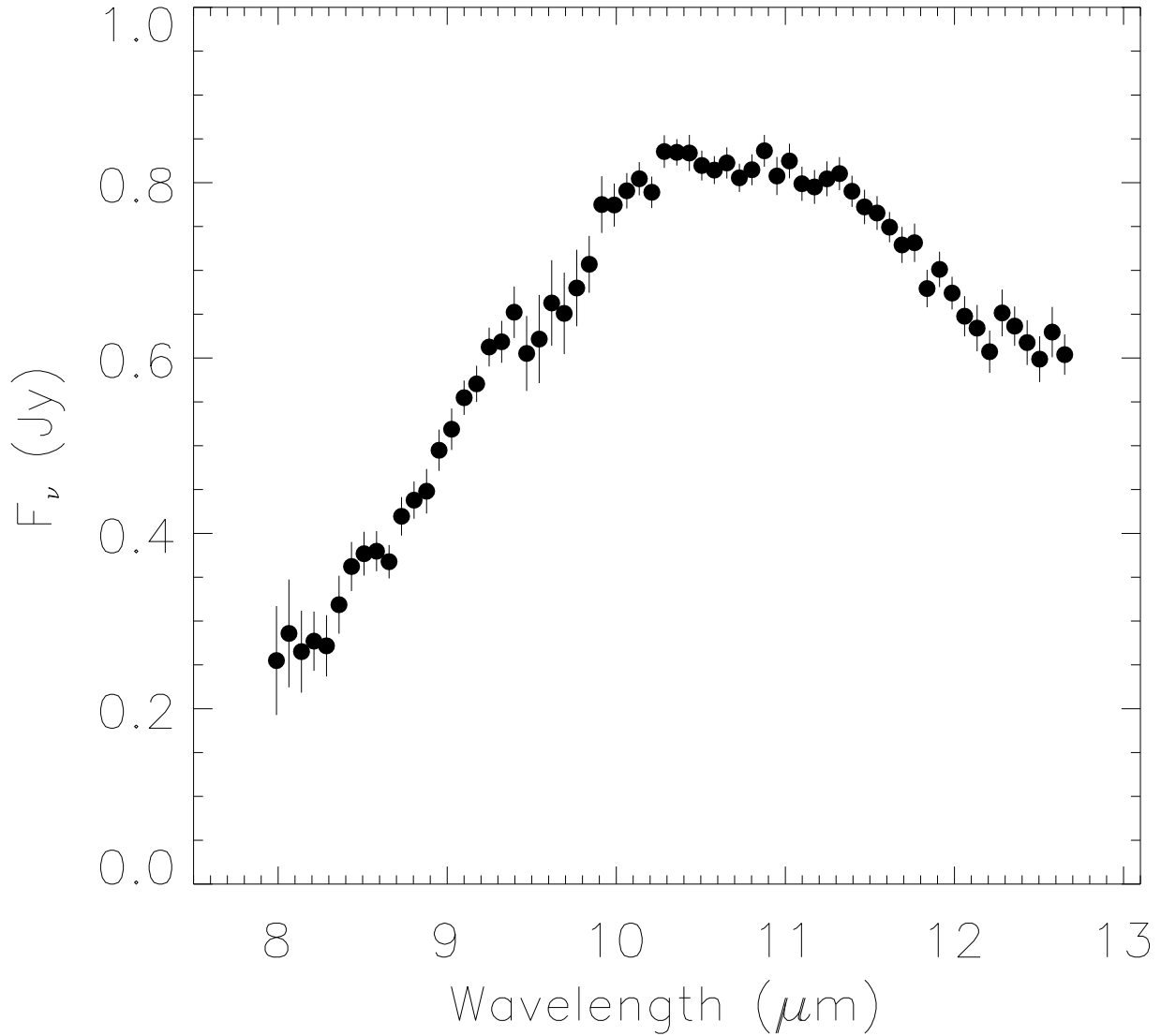} \figcaption[f6.eps]{The 8--12.5 $\mu$m spectrum of TW
Hya with resolution $\lambda / \Delta\lambda \sim 120$. The emission in
this region shows the character of amorphous silicates (peaking at
$\sim$9.6\micron) mixed with crystalline silicates (with peaks at
$\sim$11.2\micron). Around 9.7$\mu$m, the small depression in flux is
not real but is caused by imperfect calibration of the time variable
atmospheric Ozone emission. The overall normalization is set to make the
11.7 $\mu$m flux equal that measured in the imaging.  The overall shape
agrees well with that observed by \citet{Sitko} but the evidence for
crystalline components is even more compelling in the flatness of the
spectrum longward of 10\micron. \label{fig_spectrum}}
\end{figure}

The 8--13$\mu$m spectrum of TW Hya is shown in Figure
\ref{fig_spectrum}.  The 11.7\micron\ flux density, above, was used to
normalize the spectrum, since the full width of that filter is contained
in the spectral range.  There is a broad hump in this region that,
similarly to \citet{Sitko}, we attribute to emission by a combination of
amorphous and crystalline silicates.

\subsection{Detection Limits for Companions}

Multiplicity in the TW Hya association as a whole is $\gtrsim$50\%,
including stellar and substellar objects \citep{Zuckerman01}, although
TW Hya itself has no known companion.  At the young age of TW Hya any
substellar companions would be quite bright in the infrared, so our
NICMOS images provide a sensitive way to search for such objects.

\begin{figure}
\plotone{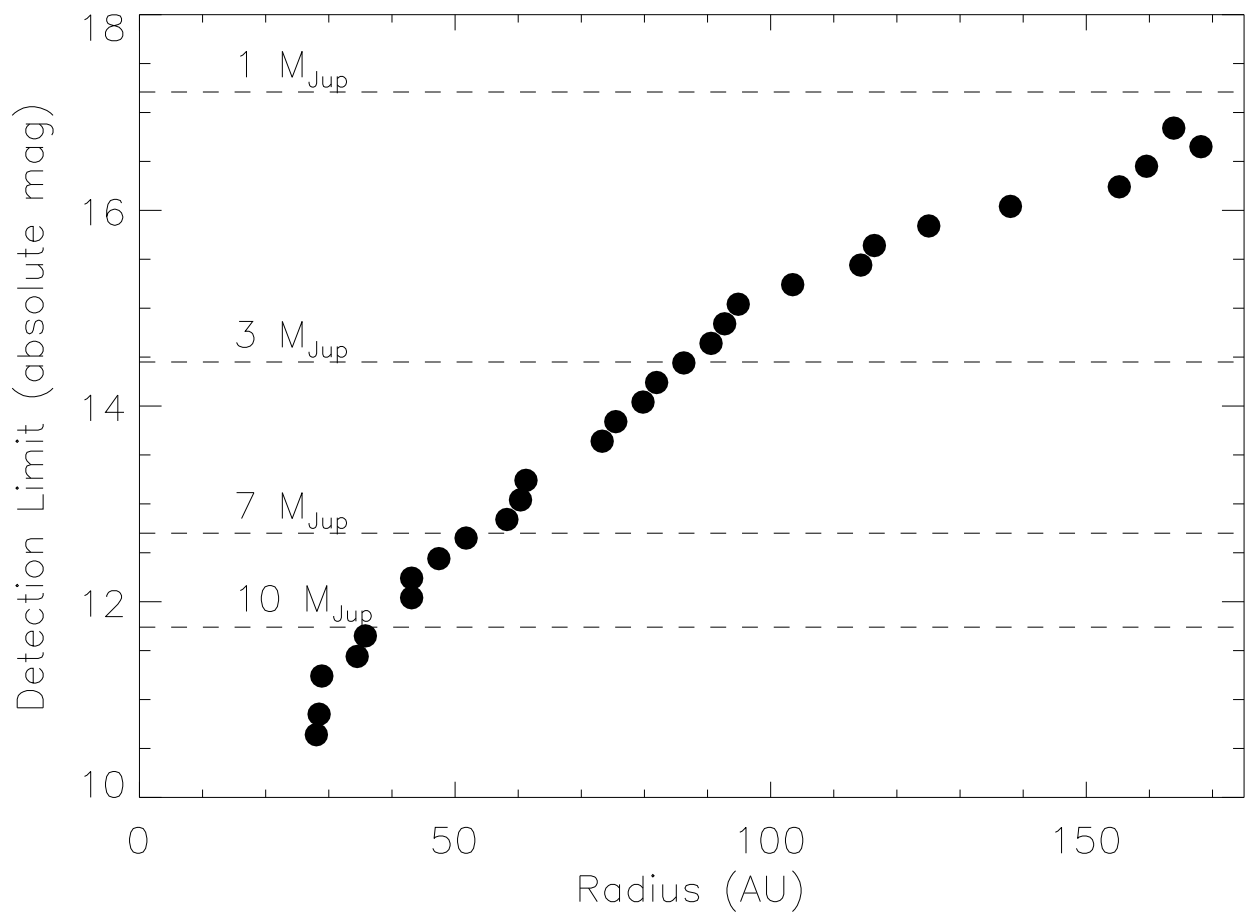} \figcaption[f7.eps]{The limiting F160W absolute
magnitude of a detectable point source as a function of distance from TW
Hya. The masses as a function of absolute magnitude are based on an age
of 10 Myr and cooling models of \protect{\citet{Burrows97}} and are
accurate to $\sim$0.3 mag (A. Burrows, personal communication).  At the
radius of $\sim$85 AU where the dip in disk surface brightness occurs,
the limit is $\sim$3M$_{Jup}$. No point source was detected within 4$''$
of TW Hya.\label{fig_ptsrclimit}}
\end{figure}

We subtracted the two orientations of our F160W data.  While the stellar
PSF, the instrumental scattering function, and detector artifacts rotate
with the aperture, any real features in the unocculted area of the
detector are unaffected by a change in the telescope/camera orientation.
Subtraction of the these two images has been shown to significantly
reduce residual PSF background light \citep{schneider98}.  Furthermore,
since the TW Hya disk is almost face-on, it nulls out nearly perfectly
in the roll-subtraction, leaving higher noise but no excess flux. In our
data, the roll was only 7\degr, so point sources in the roll subtraction
are separated by more than their FWHM for radii greater than 1.2\arcsec.
For smaller radii, the roll-subtraction was not useful for point source
detection and we used the PSF subtracted images discussed already.  In
both regions, we assessed the observability of point sources by planting
Tiny Tim PSFs and measuring the magnitude that could be recovered at
$>$3$\sigma$ significance.  No point sources were found within
3.4\arcsec\ of TW Hya.  The F160W detection limits are shown in Figure
\ref{fig_ptsrclimit}.

\section{Discussion}

\subsection{Disk Morphology} \label{diskmorph}

Given the large fraction (0.25) of the stellar luminosity re-radiated
in the mid and far-infrared, the disk must be optically thick.  For
single scattering by a geometrically flat and optically thick disk, the
flux density per square arcsec would be proportional to r$^{-3}$
\citep{whitney92}.  This is somewhat steeper than the r$^{-2.6}$ we fit
in the 40--150 AU annular region and suggests that the disk is instead
flared. Perhaps the strongest argument in favor of disk flaring is that
the reflected light surface brightness throughout this annulus is much
larger than predicted by a flat disk model.

A warm optically thick disk composed of well mixed dust and gas will
naturally flare as a consequence of vertical hydrostatic equilibrium
\citep{kenyon87}.  If dust and gas are well mixed and in interstellar
proportions, then the height at which stellar photons are scattered is
roughly proportional to the vertical pressure scale height.  In this
case, the scattering height scales as $r^{\gamma}$, where $\gamma
\approx 1.2$--$1.4$ \citep{Chiang,DAlessio98,Chiang01}.  Such a
well-mixed, flared disk has a scattering brightness profile that scales
approximately as $r^{-2}$ \citep{whitney92}, significantly
shallower than the average TW Hya profile at disk radii greater than 40
AU. The discrepancy could reflect vertical settling of dust with respect
to the gas such that the actual scattering height at these radii is
characterized by $\gamma < 1$.

\begin{figure}
\plotone{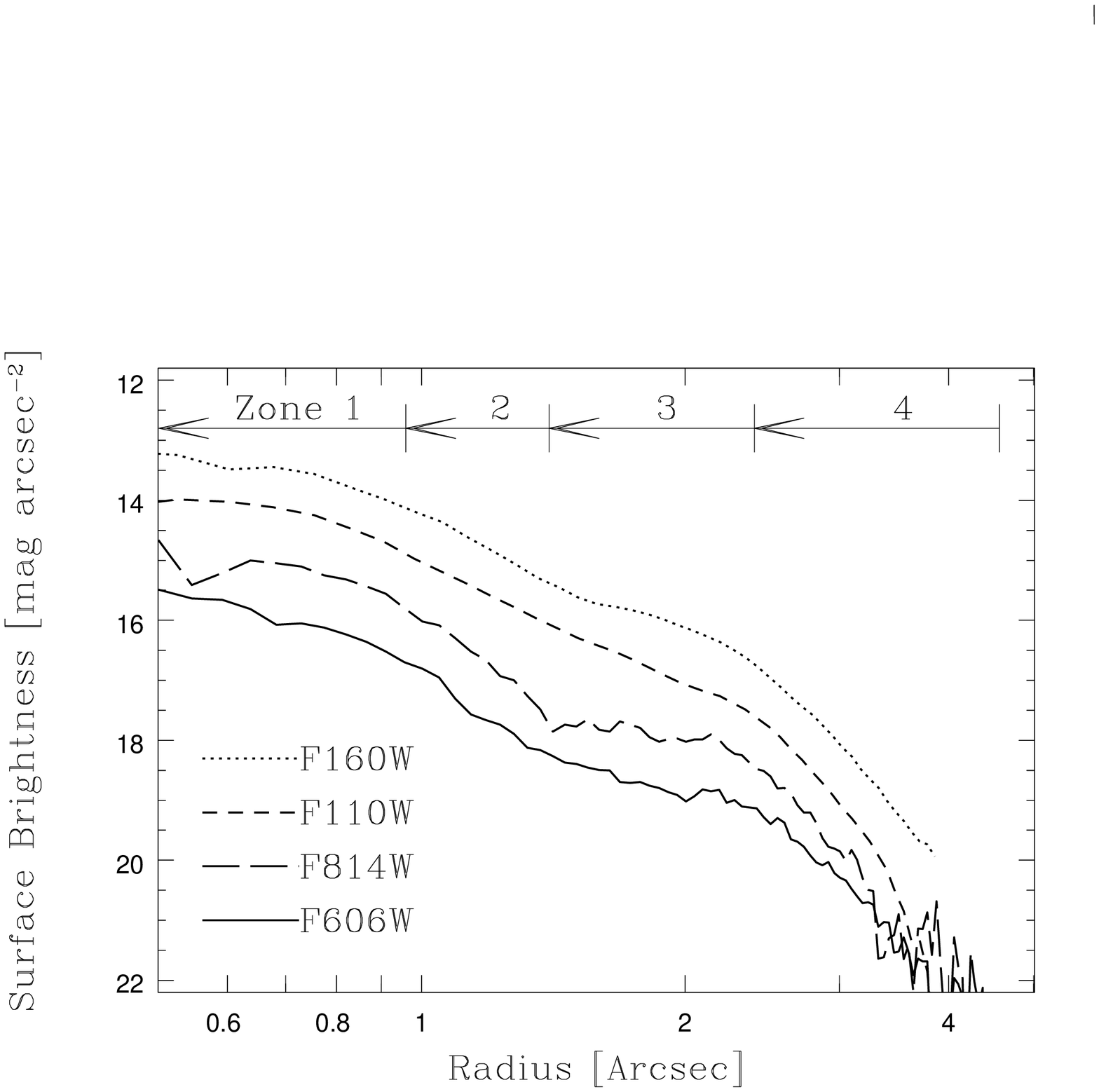}
\figcaption[f8.eps]{Profile of TW Hya disk observed with NICMOS
compared to that observed with WFPC2 by \citet{Krist00}.  Both sets of
observations show a small depression in surface brightness at 1.$''$5 as
well as the same overall slope of the surface brightness.
\label{fig_compare}}
\end{figure}

The F160W surface brightness profile between 80 and 130 AU scales like
r$^{-2}$.  This annulus corresponds to the ``zone 3'' defined by
\citet{Krist00} (Figure \ref{fig_compare}), where the visual scattered
light also flattens out in their images. The appearance of the same
feature in data from two different instruments lends credibility to the
idea that this feature is real.  Such a variation in the slope of the
surface brightness profile might reflect an undulation or ripple in the
height of the disk surface; at 40 AU $\lesssim$ r $\lesssim$ 80 AU, the
disk surface might be concave down, while at 80 AU $\lesssim$ r
$\lesssim$ 130 AU, the disk might be concave up.  Distortions of the
disk may be caused by instabilities afflicting passively heated disks
\citep{Chiang00,Dullemond00} or by dynamical effects created by a
massive body orbiting in the disk.  The detection limit on a massive
body at 100 AU from the star is 3M$_{Jup}$.  Beyond 150 AU, the surface
brightness profile is significantly steeper than r$^{-3}$.  The outer
disk is probably completely shadowed.

The face-on geometry of the disk and the average NICMOS data power law
for the disk surface brightness are consistent with the WFPC2 data of
\citet{Krist00}. Their average power law in the same 40 -- 150 AU region
(see Figure \ref{fig_compare}) is -2.4 at F606W (0.6 \micron) and -2.6
at F814W (0.8 \micron).  In addition, the geometry of the disk seen in
both HST data sets is consistent with the general morphology reported by
\citet{Trilling01} from ground-based observations.  However, there are
notable differences between the NICMOS and Trilling et al. data.  At
their innermost detectable radius of 0.$''$94, they report a peak
surface brightness at H-band of 6.8$\pm$1.1 mJy arcsec$^{-2}$.  At this
same radius, which is well outside our coronagraph, and using the F160W
filter which is very similar to H-band, we measure 2.5 $\pm$ 0.4 mJy
arcsec$^{-2}$.  In addition, they measured a total disk flux density in
the 0.$''$94--4$''$ region of 21.6 $\pm$ 3.5 mJy, whereas we measure
12.0 $\pm$ 1.2 mJy in that same region.  The Trilling et al. surface
brightness power law index, -3.3 $\pm$ 0.3, is discrepant by 2.5
$\sigma$ with the NICMOS index of -2.6 $\pm$ 0.1.  Systematic
uncertainties in their ground-based PSF subtraction under conditions of
0.$''$8 seeing may be responsible for the discrepancies.

The large optical depth of the disk makes it difficult to use the
scattered light to estimate the density of grains.  At any given
position in the disk, visible and near-infrared stellar radiation
penetrates to an optical depth of one.  In contrast, the disk is
presumably optically thin at millimeter wavelengths, so millimeter
images of the dust emission trace the true surface density.


\subsection{Composition}

To elucidate the dust composition, we combine information from the
spectral energy distribution, mid-infrared imaging, and mid-infrared
spectroscopy.

\subsubsection{SED Fitting compared to Mid-infrared Imaging}\label{midirsection}

A simple model of the disk using grains that absorb and emit like
blackbodies in thermal equilibrium with a star of luminosity 0.3
L$_{\odot}$ predicts a temperature of 214 K, the 11.7 to 17.9 \micron\
color temperature, at a distance of only 0.9 AU (0.$''$04). If the
entire disk could be described by such grains, it would therefore not be
surprising that it appears point-like.  In more sophisticated models of
passive circumstellar disks, \citet{Chiang} and \citet{Chiang01}
predict the mid-infrared flux emitted by a disk as a function
of radius.  In their two-layer disk models, the hotter disk surface
layer is heated by direct exposure to starlight, while the cooler disk
interior is heated by radiation from the surface. To compare predictions
of their models with our mid-infrared images, we modeled TW Hya after
the prescription of \citet{Chiang01} which accounts explicitly for grain
size distributions and laboratory-based silicate and water ice
opacities. Our procedure was first to select disk parameters so as to
reproduce the observed SED only.  In fitting the disk parameters, no
regard is given to any imaging data. Only after a disk model is chosen
that provides a satisfactory fit to the SED do we derive surface
brightness profiles at 11.7 \micron\ and 17.9 \micron\ based on our
fitted model disk.  These profiles are convolved with our respective
PSFs and compared to our images of TW Hya. 

Figure \ref{fig_sed} displays the observed and fitted SEDs of TW
Hya. Model input parameters are given in Table \ref{tab_irmodel}. For
additional details of the model, see \citet{Chiang01}.  The SED model
predicts that 99.8\% of the 12 $\mu$m flux density and 94.6\% of the 18
$\mu$m flux density arise within a radius of 9 AU (2 pixels) of the
star.  As is shown in Figure \ref{fig_midirimages}, at the resolution of
the Keck Telescope, no extended emission is predicted by the model. This
is consistent with the observations in which no extension is observed.

\begin{deluxetable}{ll}
\tablecaption{Parameters of TW Hya Disk Model\label{tab_irmodel}}
\tablewidth{0pt}
\tablehead{
\colhead{Parameter} &\colhead{Value}
}
\startdata
maximum grain radius in surface   &1 $\mu$m \\
maximum grain radius in interior  &6000 $\mu$m \\
grain core size distribution ($dN/dr$) &\\
{\hskip0.3in}in disk interior          &$\propto r^{-1}$ \\
{\hskip0.3in}in disk surface           &$\propto r^{-3.5}$ \\
dust surface mass density         &10 (r/AU)$^{-1}$ g cm$^{-2}$ \\
inner disk radius                 &0.05 AU \\
outer disk radius                 &200 AU \\
\enddata
\end{deluxetable}

Outside 0.2 (2) AU, silicate cores are mantled by water ice in the disk
interior (surface).  The remarkably large grains in the disk interior,
consisting of millimeter--centimeter sized cores, and the large total
condensable mass in the disk of 0.0014 M$_\odot$ = 470 M$_\oplus$, were
necessary to fit the millimeter-wave SED measured by \citet{Weintraub89}
and \citet{Wilner00}.  The measurement at the particularly long
wavelength of 7~mm constrains the grain size distribution in the disk
interior to be significantly flatter than the standard interstellar law
of $dN/dr \propto r^{-3.5}$, and to possess an upper size cut-off of
6~mm.  Had we employed a grain size distribution that was much steeper
than $r^{-1}$ and that had an upper size cut-off smaller than $r=$~6~mm,
the model fluxes between $1.3$ and 7~mm would have been too low
compared to the measurements. Our fitted grain size distributions in the
disk interior are subject to the well-known degeneracy between grain
size and disk surface density (see, e.g., \citet{Chiang01});
reducing the grain sizes or steepening the size distribution increases
the fitted dust surface mass density. However, increasing the fitted
dust surface density would threaten to make the disk gravitationally
unstable, if gas and dust are present in interstellar proportions in TW
Hya. We draw the model-dependent conclusion that grain growth has indeed
occurred in the TW Hya disk.

For the particular grain mixture of millimeter to centimeter-sized
olivine spheres mantled with water ice employed by our model,
$\kappa_{1.1 {\rm mm}, \rm\,dust} = 0.84 \rm\,cm^2 \rm\,g^{-1}$ and
$\kappa_{7 {\rm mm}, \rm\,dust} = 0.14 \rm\,cm^2 \rm\,g^{-1}$.  A
somewhat lower dust mass of $\sim$130 $M_{\oplus}$ was derived by
\citet{Wilner00} and \citet{Trilling01} using the same observational
data.  The difference arises largely because their assumed dust
opacities were $\sim$3 times greater than ours (we have multiplied their
opacities by 100 to isolate the dust component). Their values were based
on the \citet{Hildebrand} measurement of $\kappa_{250 \mu{\rm m},
\rm\,dust}$ = 10 cm$^2$ g$^{-1}$ in the dark molecular cloud NGC 7023
and extrapolated to longer wavelengths with $\kappa_{\nu, \rm\,dust}
\propto \nu^{\beta}$, where $\beta$ is taken to be 1.0 by Wilner et
al. and 0.9 by Trilling et al.  The assumed similarity between dust in
NGC 7023 and dust in TW Hya has no direct observational evidence.

\citet{Pollack94} showed that a single power law description of the
opacity is inaccurate for real dust materials.  They computed opacities
for a variety of grain species, including water ice, silicates,
organics, iron, and troilite. For their ``Composite (50\%)'' grains,
extrapolated values of $\kappa_{7 {\rm mm}, \rm\,dust}$ range between
0.03 and 0.06 cm$^2$ g$^{-1}$, assuming that the values of $\beta$ that
they compute between 650 $\mu$m and 2.3 mm can be extended to 7
mm. These opacities are a factor of $\sim$3 smaller than the ones
employed in our disk model fit, and a factor of 5--10 smaller than the
ones adopted by \citet{Wilner00} and \citet{Trilling01}.

\citet{Krist00} model only $\sim$33 $M_{\oplus}$ of ISM-like dust to
reproduce their observed surface brightness profiles in scattered light
at $\lambda$ = 0.6 and 0.8 $\mu$m.  The profiles depend more strongly on
the distribution of dust at high altitude above the midplane than on the
distribution of dust in the midplane where most of the mass probably
resides. Krist et al. assume a Gaussian vertical density profile for
micron-sized dust.  However, substantially more dust could reside at the
midplane of their disk model without detracting from the goodness of
their surface brightness fits.  Thus, their $\sim$33 $M_{\oplus}$ could
easily represent a lower limit on the actual dust mass.

We conclude that the TW Hya disk probably contains several hundred Earth
masses of condensed silicates and ices. This dust mass is several times
larger than the typical dust masses estimated for T Tauri star disks in
the Taurus and Ophiuchus star forming regions on the basis of 1.3 mm
dust continuum observations \citep{Osterloh95,Andre94}.  Large,
centimeter sized, grains were necessary to fit the $\sim\nu^{3.5}$ slope
of the millimeter-wave SED.  A grain size distribution with smaller
grains is characterized by a significantly steeper slope,
$\sim\nu^{4-5}$, and is difficult to reconcile with the observations.
The presence of these very large particles is good evidence for grain
growth within the disk.

Total mass estimates for TW Hya's disk in the literature vary by another
two orders of magnitude depending on the assumed gas:dust ratio.  The
standard interstellar gas:dust mass ratio of 100:1 may be incorrect for
an older disk like that around TW Hya.  \citet{Kastner97} find only 11
M$_\oplus$ in H$_2$ gas based on $^{13}$CO observations and a standard
interstellar H$_2$/CO ratio.  The low mass accretion rate estimated by
\citet{Muzerolle}, an order of magnitude less than for typical T Tauri
stars, also suggests that the gas in this system has been depleted.  The
gas-to-dust mass ratio does not formally enter into our model SED-fitted
disk; only enough gas and/or vertical circulation of gas is assumed to
lift dust grains $\sim$3--5 vertical scale heights above the midplane to
comprise the flared, superheated disk atmosphere.  If we assume the
primordial TW Hya disk had the canonical 100:1 gas to dust ratio, then
it once contained 0.14 M$_\odot$.  This early disk with $\sim$25\% of
the star's mass would have been marginally gravitationally stable.

\subsubsection{Mid-infrared spectrum}

The broad silicate emission feature cannot be explained by amorphous
silicates alone since they have a pronounced peak at 9.6\micron\ and
fall off rapidly to longer wavelengths.  Our spectrum agrees very well
with the published result of \citet{Sitko}, who note the near absence of
a peak at 11.2 $\mu$m, within the uncertainties of their data.  Our
spectrum hints at a peak at 11.2 $\mu$m; furthermore, to create the full
width of the feature requires substantial flux longward of
10\micron. Crystalline species are probably responsible.  Data of this
spectral resolution are inadequate to constrain the exact crystalline
structure, however.  Magnesium and iron-rich species all have peaks in
the 10 -- 11.3\micron\ region \protect{\citep{Jager98}}.  Silicate
emission features become less prominent as the grains get larger, so
that by a size of $\sim$5\micron\ their mid-infrared spectrum is nearly
flat \citep{skinner92}.  The TW Hya spectrum suggests that its emitting
grains must therefore be $\lesssim$5$\mu$m in size.

\section{Conclusions}

TW Hya is surrounded by a optically thick dust disk which must be flared
to account for the large apparent size of the disk seen in scattered
light and for its thermal SED.  A ripple in the surface brightness at 85
AU and the steep decline in surface brightness beyond 150 AU could be
due to thermal instabilities or dynamical effects.  However, no
companion to TW Hya is detected down to a mass of $\sim$10 M$_{Jup}$ at
distances greater than 50 AU from the star.

Although TW Hya looks like a classical T Tauri star in terms of its
H$\alpha$ emission, its disk shows signs of evolution.  To fit the large
submillimeter and millimeter flux densities measured from TW Hya, our
SED models require a grain size distribution in the disk interior
dominated by large, millimeter-centimeter sized, icy grains.  Large
grains, which dominate the mass of the disk, presumably grew from
coagulation of smaller grains in the disk.  The initial mass of TW Hya's
disk in gas and dust may be several times the typical T Tauri star disk
mass, which may help account for its long, $\sim$8 Myr, lifetime.  The
thermal emission remains spatially unresolved at the resolution of the
Keck Telescope, in accord with theoretical expectations.

\acknowledgements

We thank John Krist and Mike Sitko for providing their data in digital
form, our anonymous referee for insightful comments on the manuscript,
and the Keck staff for help with LWS.  This work was supported by NASA
grant NAG 5-3042 to the NICMOS IDT and by a Hubble Fellowship grant to
EIC. This paper is based on observations with the NASA/ESA Hubble Space
Telescope, obtained at the Space Telescope Science Institute, which is
operated by the Association of Universities for Research in Astronomy,
Inc. under NASA contract NAS5-26555.  This paper is also based on
observations at the W. M. Keck Observatory, which is operated as a
scientific partnership between the California Institute of Technology,
the University of California, and the National Aeronautics and Space
Administration and was made possible by the generous financial support
of the W. M. Keck Foundation.

\end{document}